# Spin Transition in the ν=8/3 Fractional Quantum Hall Effect

W. Pan[1], K.W. Baldwin[2], K.W. West[2], L.N. Pfeiffer[2], and D.C. Tsui[2]

[1] Sandia National Laboratories, Albuquerque, New Mexico, USA 87185
[2] Princeton University, Princeton, New Jersey, USA 08544

Abstract

We present here the results from a density dependent study of the activation energy gaps of the fractional quantum Hall effect states at Landau level fillings ν=8/3 and 7/3 in a series of high quality quantum wells. In the density range from $0.5 \times 10^{11}$ to $3 \times 10^{11}$ cm$^{-2}$, the 7/3 energy gap increases monotonically with increasing density, supporting its ground state being spin polarized. For the 8/3 state, however, its energy gap first decreases with increasing density, almost vanishes at n ~ $0.8 \times 10^{11}$ cm$^{-2}$, and then turns around and increases with increasing density, clearly demonstrating a spin transition.

The fractional quantum Hall effect (FQHE) [1,2] in the second Landau level has attracted a great deal of interests in recent years due to its possible applications in fault-resistant topological quantum computation [3]. Tremendous advance has been achieved in understanding the most celebrated 5/2 FQHE state, believed to be due to paring [4] of composite fermions (CF) [5-7] and that its elementary excitations obey non-Abelian statistics.

In addition to the 5/2 state, many odd-denominator FQHE states have also been observed, for example at Landau level fillings $\nu=7/3$ and 8/3 [8-21]. In contrast to the 5/2 state, much less work has been carried out for these states. On the other hand, unlike the odd-denominator FQHE state in the first Landau level, where most of them are well understood within the picture of either the hierarchical model [22,23] or CF model [5-7], the nature of the odd-denominator FQHE states in the second Landau level remains largely unsettled [24]. This is even true for the most prominent ones at the simplest odd-denominator Landau level fillings $\nu=7/3$ and 8/3. Indeed, a Laughlin type FQHE state was originally ruled out for these two states based on finite size, few particles calculations [25,26]. More recent detailed calculations have also shown that the model of weakly interacting composite fermions is not adequate for these second Landau level fractions [24]. Over the years, proposals of novel ground states [27-37] have been put forward. It is expected that a deep understanding of the FQHE in the second Landau level will lead to much exciting many-body physics [24].

Experimentally, currently available transport results appear more complex than expected from a simple analogy of their counterparts (the $\nu=1/3$ and 2/3 FQHE states) in the first Landau level. For example, it has been observed by many groups that the energy gap of the 7/3 state is roughly two times that of the 8/3 state. This difference cannot be explained by assuming these two states are particle-hole conjugate states and, thus, by the slight difference in *B*-field at $\nu=7/3$ and $\nu=8/3$. As a result, an explanation related to spin polarization was proposed [13]. Naively, extrapolating from the lowest Landau level, one might expect that the 7/3 state is spin polarized, whereas the 8/3 state is unpolarized.

However, one earlier theoretical paper [38] predicts that the ν=8/3 state is also spin-polarized even at vanishingly small Zeeman energies.

To study the spin-polarization of a FQHE state, the commonly used experimental technique is to tilt sample in-situ in magnetic fields at very low temperatures [39-41]. By so doing, one varies the relative strength of the Zeeman energy ($E_z$) and the Coulomb energy ($E_c$), where $E_z = g^*\mu_B B_{total}$ and $E_c = e^2/\varepsilon l_B$. $g^*$=0.44 is the effective g-factor, $\mu_B$ the Bohr magneton. $B_{total} = B_{perp}/\cos(\theta)$ is the total magnetic field under tilt, $B_{perp}$ the perpendicular magnetic field to the sample normal and θ the tilt angle. $l_B = (\hbar/eB_{perp})^{1/2}$ is the magnetic length, ℏ the Planck constant, e the electron charge. ε is the dielectric constant of GaAs. However, this technique appears to be complicated to tackle the spin polarization in the second Landau level due to a strong coupling of the orbital motion. Indeed, experimental attempts [42-47] under this approach have shown surprisingly complex behaviors. First, it was observed [42,43] that the in-plane magnetic field from tilting can induce a phase transition from the quantum Hall effect phase to an anisotropic phase in the second Landau level. Then, the mixing of different electric subbands under tilt can give rise to totally different tilt magnetic field dependence of the 7/3 and 8/3 energy gaps in samples of different well width [47], thus making asserting their spin polarization almost impossible.

In this paper, we use a different approach and study the spin polarization of the 7/3 and 8/3 states as a function of electron density (n). Under this approach, the B-field is always perpendicular to the two-dimensional electron system (2DES). By changing the 2DES density, the ratio of Coulomb energy $E_c$ to the Zeeman energy $E_z$ also changes, since $E_c \sim n^{1/2}$ and $E_z \sim n$. In this regard, the density dependence approach is equivalent to tilting magnetic field but it cannot cause a tilt-field induced phase transition. It is observed that in the density range between $0.5 \times 10^{11}$ and $3 \times 10^{11}$ cm$^{-2}$, the energy gap of the 8/3 state ($\Delta_{8/3}$) first decreases with increasing density, nearly disappears at n ~ $0.8 \times 10^{11}$ cm$^{-2}$. Beyond this density, $\Delta_{8/3}$ increases with increasing density. This density dependence of $\Delta_{8/3}$ clearly signals a spin transition at this filling factor. For comparison, the energy gap of the 7/3 state ($\Delta_{7/3}$) shows a monotonic density dependence, supporting a spin polarized state down to $0.5\times10^{11}$ cm$^{-2}$.

The specimens we used in this study are a series of high quality symmetrically doped GaAs quantum wells [48]. Table I lists the sample parameters, including the 2DES density, mobility, and quantum well width (W), and the ratio of W/$l_B$ at the Landau level filling ν=8/3. The low-temperature electron density and mobility were established by a brief red light-emitting diode illumination at 4.2K. Standard low-frequency lock-in technique (~ 11Hz) was utilized to measure the magnetoresistance $R_{xx}$ and Hall resistance $R_{xy}$.

In Figure 1a, we show the $R_{xx}$ trace for sample C. A fully developed 5/2 state is clearly seen at B ~ 1.3T, i.e., vanishingly small $R_{xx}$ and a quantized $R_{xy}$ (not shown). This is so far the lowest B field that a fully developed 5/2 FQHE state has been reported. $R_{xx}$ minimum is also observed at other filling factors ν=7/3, 8/3, 11/5, and 14/5. In Fig.1b, a semi-log plot of $R_{xx}$ versus 1/T is shown for ν=8/3 and 7/3. From fitting, the energy gaps at these two fillings are obtained: $\Delta_{7/3}$ ~ 35 mK and $\Delta_{8/3}$ ~ 10 mK.

In Fig. 1c, we show the $R_{xx}$ trace at a lower electron density of n=$0.5\times10^{11}$ cm$^{-2}$. In this lower density sample, only the strongest FQHE states at ν=8/3, 5/2, and 7/3 are seen. What is really surprising is that the 8/3 state is the strongest among the three FQHE states. This is also corroborated when examining their activation energy gaps (shown in Fig.1d): $\Delta_{7/3}$ ~ 5 mK and $\Delta_{8/3}$ ~ 45 mK.

In Figure 2a and 2b, we plot the energy gaps at ν=8/3 and 7/3 as a function of electron density. It is clear that the energy gap of the 8/3 state first decreases with increasing density, nearly disappears at n ~ $0.8\times10^{11}$ cm$^{-2}$. Beyond this density, $\Delta_{8/3}$ increases with increasing density. This change observed in the 8/3 energy gap is very similar to what was observed in the ν=2/3 FQHE in the lowest Landau level [49,50] and demonstrates a spin transition [49-57] from a spin unpolarized ground state at low densities to a spin polarized one at higher densities. For comparison, $\Delta_{7/3}$ shows a monotonic density dependence, supporting that the 7/3 state is spin-polarized down to $0.5\times10^{11}$ cm$^{-2}$.

Before we discuss the implications of the above observation, we want to point out that the observed spin transition is intrinsic and cannot be induced by extrinsic means, such as finite thickness [58] or Landau level mixing [59]. First, it has been shown that the spin polarization of a FQHE state is insensitive to the finite-thickness correction [38]. Second, in this experiment, the quantum well width is varied in accordance with the electron density so that the parameter, $W/l_B$, a measure of effective thickness of 2DES, remains more or less the same in all samples, as shown in Table I. Consequently, the percentage of the reduction to the energy gap calculated for an ideal 2DEG is roughly the same for all the samples. The Landau level mixing (LLM) effect cannot cause the above spin transition, either. It is known that LLM is strong at low electron densities [59]. As a result, the reduction of energy gap due to LLM should be larger at low densities, actually smearing the sharpness of transition if the intrinsic gap were plotted.

In a recent publication, Liu et al showed there exists a giant enhancement in the 5/2 energy gap in the vicinity of the crossing between Landau levels belonging to the different (symmetric and antisymmetric) electric subbands [19]. A self consistent calculation for our samples has ruled out this possibility for a large ν=8/3 energy gap in the low density regime.

The observation of a spin transition at 8/3 is contradictory to the conclusion reached in Ref. [38], where the authors found from their numerical calculation that the 8/3 state was different from the 2/3 state and remained spin polarized even at vanishingly small Zeeman energy. This is, as they argued, because the more repulsive effective interactions in the second Landau level force electrons to occupy the maximum spin state. Our experimental results, however, show that the 8/3 state behaves very much like the 2/3 state and display a spin transition as a function of density. One may argue that the theoretical calculation was carried out at a 2DES density of ~ $2.8\times10^{11}$ cm$^{-2}$, which is much larger than the transition density of $0.8\times10^{11}$ cm$^{-2}$. On the other hand, the relevant parameter in determining the spin polarization of a FQHE state is the ratio of the Zeeman energy $E_z$ to Coulomb energy $E_c$ [60]. At $n=0.5\times10^{11}$ cm$^{-2}$, $E_z/E_c \sim 0.005$. Using the parameters quoted in Ref. [38], $n=2.8\times10^{11}$ cm$^{-2}$ and $g^*=0.05$, $E_z/E_c$ is much smaller, $\sim 0.0015$. Thus, the 8/3 state considered in Ref. [38] should be deeper in the unpolarized regime, instead of being fully polarized predicted by the theoretical calculations.

A spin unpolarized ground state at $\nu=8/3$ is also inconsistent with the models of a spin-polarized non-Abelian state for the 3$^{rd}$ FQHE states in the second Landau level. On the other hand, it remains unclear whether it can be a two-component non-Abelian state [36], or a paired spin-singlet quantum Hall state [28], or a boundary state between the Abelian and non-Abelain states [35]. Our current data are not able to address this question.

The observation of a spin transition at 8/3 and a spin polarized 7/3 state, on the other hand, is mostly consistent with the composite fermion model with a spin [61]. This can be derived from a simple analogy of their counterparts in the first Landau level. Under the CF model, the 7/3 state is mapped onto the $\nu^*=1$ interger quantum Hall effect (IQHE) state of the CFs emanating from the 1/2 state in the second Landau level and, thus, is spin polarized. The 8/3 state is the $\nu^*=2$ IQHE sate of the CFs and is spin unpolarized at small effective magnetic fields, or low electron densities. With increasing density, CF Landau level crossing can occur [61] and the 8/3 state becomes spin-polarized beyond the critical density.

One remark is in order before we conclude this paper. Unlike in the high density regime where $\Delta_{7/3}$ is roughly twice of $\Delta_{8/3}$, at $n=0.5\times10^{11}$ cm$^{-2}$ $\Delta_{7/3}$ is much smaller than $\Delta_{8/3}$. In fact, $\Delta_{8/3} \sim 10 \times \Delta_{7/3}$. This big difference probably can be explained under the CF model with a spin, where the energy gap at $\nu^*=1$ or $\nu=7/3$ is due to Zeeman splitting of CFs and the energy gap at $\nu^*=2$ or $\nu=8/3$ is due to cyclotron gap. Alternatively, it is possible that the 7/3 state may also be spin unpolarized at even lower electron densities than studied in this experiment, and the spin transition occurs very close to $0.5\times10^{11}$ cm$^{-2}$, where a tiny 7/3 gap was observed. On the other hand, a spin-unpolarized 7/3 state is not expected under the CF picture.

In summary, we have carried out density dependence of the energy gaps at ν=8/3 and 7/3 in a series of high quality quantum wells. A spin transition is observed in the 8/3 FQHE. The 7/3 state appears to be spin polarized down to $0.5 \times 10^{11}$ cm$^{-2}$.

We would like to thank J.K. Jain, Th. Jolicoeur, E. Rezayi, M. Shayegan, and A. Wójs for discussions. The work at Sandia was supported by the DOE Office of Basic Energy Science. Sandia National Laboratories is a multi-program laboratory managed and operated by Sandia Corporation, a wholly owned subsidiary of Lockheed Martin Corporation, for the U.S. Department of Energy's National Nuclear Security Administration under contract DE-AC04-94AL85000. The work at Princeton was supported by the DOE under Grant No. DE-FG02-98ER45683, and partially funded by the Gordon and Betty Moore Foundation as well as the National Science Foundation MRSEC Program through the Princeton Center for Complex Materials (DMR-0819860).

Table I. The quantum well width (W), 2DES density and mobility, as well as the magnetic length ($l_B$) at $\nu=8/3$ and the ratio of W/$l_B$ for the samples studied in this work.

| samples | well width (nm) | density ($10^{11}$ cm$^{-2}$) | mobility ($10^6$/V s) | $l_B$ at $\nu=8/3$ (nm) | W/$l_B$ |
|---|---|---|---|---|---|
| A | 60 | 0.5 | 10 | 29.2 | 2.1 |
| B | 60 | 0.6 | 9.1 | 26.7 | 2.2 |
| C | 56 | 0.77 | 13 | 23.6 | 2.4 |
| D | 45 | 1.15 | 13.8 | 19.3 | 2.3 |
| E | 33 | 2.1 | 23 | 14.3 | 2.3 |
| F | 30 | 2.6 | 24 | 12.9 | 2.3 |
| G | 30 | 3.1 | 31 | 11.8 | 2.5 |

Figures

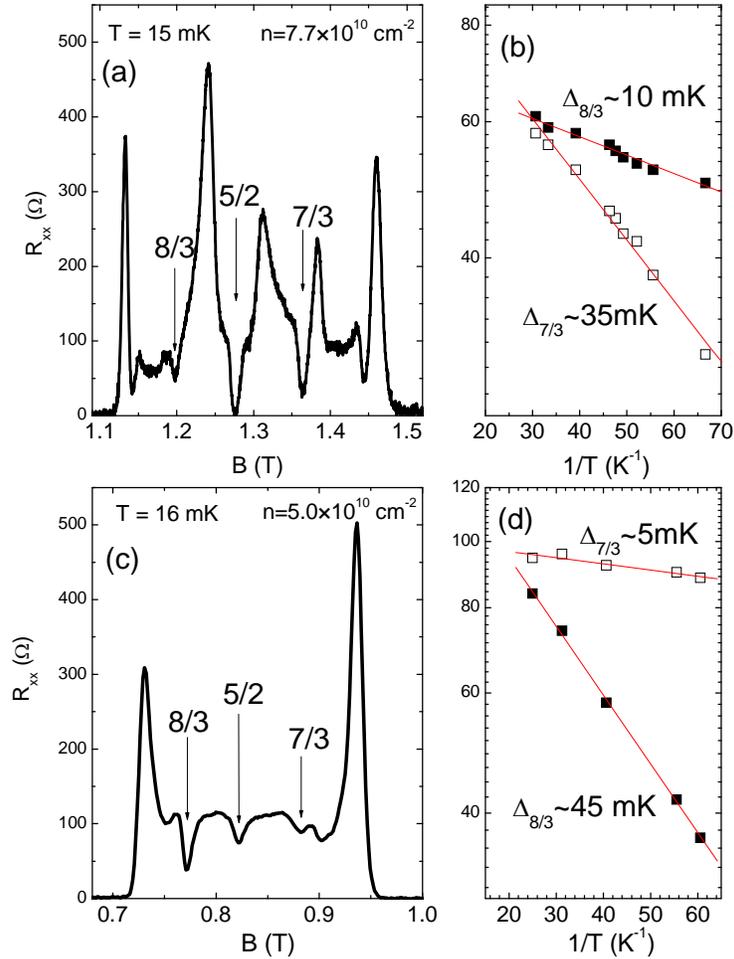

Figure 1: Magneto-resistance $R_{xx}$ for sample C (Fig. 1a) and A (Fig. 1c). Arrows mark the positions of the FQHE states at ν=8/3, 5/2, and 7/3. Fig. 1b and Fig. 1d show the temperature dependence of $R_{xx}$ at ν=8/3 (filled squares) and 7/3 (open squares) in these two samples, respectively. The lines are linear fit.

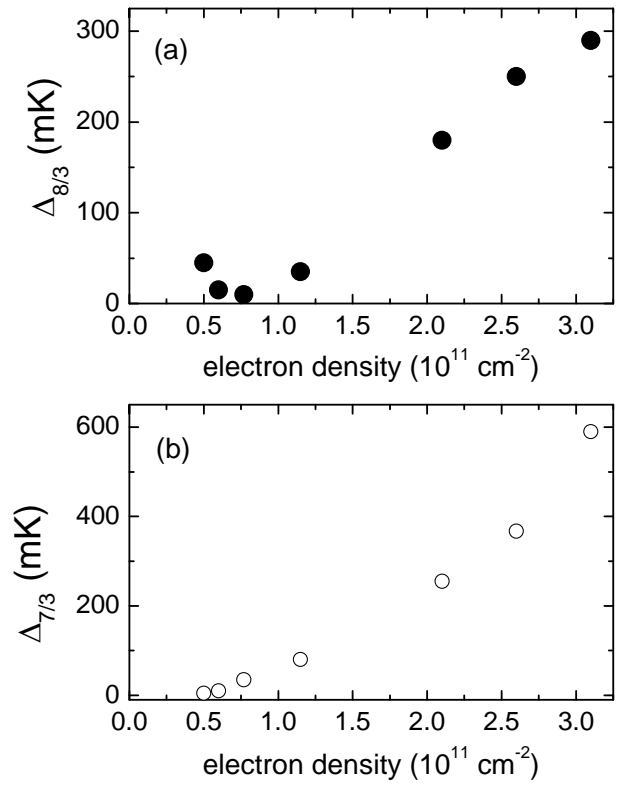

Figure 2: Activation energy gap at $\nu=8/3$ (a) and $7/3$ (b) as a function of density.